\documentclass{article}
\usepackage[a4paper, total={5in, 8in}]{geometry}
\usepackage[utf8]{inputenc}
\usepackage{quattrocento}
\usepackage{textcomp,marvosym}
\usepackage[T1]{fontenc}
\usepackage[british]{babel}
\usepackage{amsmath}
\usepackage{amsfonts}
\usepackage{amssymb}
\usepackage{graphicx}
\usepackage{caption}
\usepackage{enumerate}
\usepackage{indentfirst}
\usepackage{xcolor}
\usepackage{url}
\usepackage{microtype}
\usepackage{tabularx}
\usepackage{appendix}
\usepackage{booktabs}
\usepackage{multirow}
\usepackage[version=4]{mhchem}
\usepackage{float}
\usepackage{lipsum}
\usepackage{soul}
\usepackage{fancyhdr}
\usepackage{enumitem}
\usepackage{xurl}

\newcommand{\matr}[1]{\mathbf{#1}}
\newcommand{\ffrac}[2]{\ensuremath{\frac{\displaystyle #1}{\displaystyle #2}}}

\usepackage[backend=bibtex,style=ieee,citestyle=numeric,date=year,mincitenames=1, maxcitenames=2,safeinputenc, url=false]{biblatex}
\graphicspath{{./figures/}}
\usepackage[linktocpage=true]{hyperref}
\hypersetup{
	colorlinks=true,
	linkcolor=blue,
	citecolor=blue,
	filecolor=magenta,      
	urlcolor=cyan,
}

\begin{document}

{\Large\noindent
\textbf{What is redundant and what is not? Computational trade-offs in modelling to generate alternatives for energy infrastructure deployment}
}

\noindent 
Francesco Lombardi\textsuperscript{1 *},
Bryn Pickering\textsuperscript{2},
Stefan Pfenninger\textsuperscript{1}
\vspace{0.3cm}

\noindent \textbf{1}  TU Delft, Faculty of Technology, Policy and Management, Department of Engineering Systems and Services, Delft, Netherlands

\noindent \textbf{2}  ETH Zürich, Institute for Environmental Decisions, Department for Environmental Systems Science, Zürich, Switzerland

\vspace{0.2cm}

\noindent Corresponding author: Francesco Lombardi\textsuperscript{*} 

\vspace{0.2cm}

\noindent\textbf{Abstract}: Given the urgent need to devise credible, deep strategies for carbon neutrality, approaches for `modelling to generate alternatives' (MGA) are gaining popularity in the energy sector. Yet, MGA faces limitations when applied to state-of-the-art energy system models: the number of alternatives that can be generated is virtually infinite; no realistic computational effort can discover the complete technology and spatial option space. Here, based on our own SPORES method, a highly customisable and spatially-explicit advancement of MGA, we empirically test different search strategies -- including some adapted from other MGA approaches -- with the aim of identifying how to minimise redundant computation. With application to a model of the European power system, we show that, for a fixed number of generated alternatives, there is a clear trade-off in making use of the available computational power to unveil technology versus spatial dissimilarity across alternative system configurations. Moreover, we show that focussing on technology dissimilarity may fail to identify system configurations that appeal to real-world stakeholders, such as those in which capacity is more spread out at the local scale. Based on this evidence that no feasible alternative can be deemed redundant a priori, we propose to initially search for options in a way that balances spatial and technology dissimilarity; this can be achieved by combining the strengths of two different strategies. The resulting solution space can then be refined  based on the feedback of stakeholders. More generally, we propose the adoption of ad-hoc MGA sensitivity analyses, targeted at testing a study's central claims, as a computationally inexpensive standard to improve the quality of energy modelling analyses.

\vspace{0.2cm}

\vspace{0.2cm}
\noindent\textbf{Highlights}:
\begin{itemize}[leftmargin=*]
	\item Finding alternatives entails a trade-off between spatial and technology dissimilarity
	\item Focussing on high-level technology alternatives may leave key options unexplored
	\item Combining different search methods into a hybrid one leads to a more balanced search
\end{itemize}

\pagebreak

\section{Introduction} \label{intro}

Large-scale energy system optimisation models are increasingly used to support the urgent task of planning for the energy transition \cite{susser_model_policymaking}. Most typically, they are used to understand how to deploy new energy infrastructure to make energy systems fully carbon-neutral at the country or continental scale while keeping the economic cost for society as low as possible \cite{deng_review_2020}. However, as reported by real-world stakeholders \cite{susser_user_needs_complexity}, and increasingly acknowledged in the literature \cite{decarolis_formalizing_2017,yue_review_2018}, the provision of a single solution that minimises total economic cost is of little use in practice, for a number of reasons. First, real-world decisions on the energy transition involve a multitude of stakeholders, including local communities, with many other concerns than the total economic cost \cite{price_implications_2020}. Second, modelled costs for future systems are uncertain, for instance due to technology cost projections that need to be best-guessed when optimising for a long-term horizon \cite{ellenbeck_how_2019}. It is thus problematic to concentrate on that configuration which ensures the minimum economic cost when some of the apparently more costly options might end up being just as or even more cost-effective after cost uncertainty is realised in practice. Third and final, the generation and communication of only a single, least-cost solution can create confusion between what is least-cost and what is possible, with dangerous consequences. For instance, it is common to hear claims that a given investment decision, say installing large amounts of bioenergy power supply, is `required' for a country's energy transition because it is featured in the least-cost solution \cite{brown_response_2018}. In practice, however, strategies without bioenergy may exist within the cost uncertainty range of the least-cost solution. 

As a solution to the pitfalls of economic optimisation, some have proposed introducing secondary objectives, such as the minimisation of CO$_2$ emissions \cite{prina_multi-objective_2020,pickering_quantifying_2021}. Yet, the alternatives obtained along a multi-objective Pareto front and its near-optimal region cannot ensure that the full range of possibilities gets captured. Other alternatives, driven by unmodelled (or even impossible-to-model) objectives, might also exist and be relevant for real-world discussion \cite{brill_use_1979}. To address the impossibility to model all that might matter in reality, often referred to as the structural uncertainty of the modelling process \cite{yue_review_2018}, approaches known as `modelling to generate alternatives', or MGA, have come to the forefront in recent years \cite{decarolis_mga_2011,price_modelling_2017,li_investment_2017,berntsen_ensuring_2017}. Conceptualised by \citeauthor{brill_use_1979} in 1979 \cite{brill_use_1979} and first applied to energy system models a decade ago \cite{decarolis_mga_2011}, the basic idea behind MGA is: first, to compute the least-cost solution as a starting point; and second, to change the objective of the problem to search for something as different as possible from the previously-found solution, while enforcing that cost does not increase too much compared to the minimum feasible system cost. The process (also known as the `Hop, Skip and Jump' algorithm) can be repeated indefinitely, each time updating the objective to search for something different from all the previously found feasible solutions. 

Yet, MGA displays limitations when applied to sufficiently large models, such as state-of-the-art energy system optimisation models covering the whole of Europe at high temporal and spatial resolution and with many technological options \cite{pickering_diversity_2022,victoria_speed_2022}. The number of meaningful alternatives that can be generated for such models is \textit{de facto} infinite, and the conventional form of applying MGA often fails to represent the range of available options well, for two reasons. 

First, the `Hop, Skip and Jump' algorithm does not span the solution space evenly, leaving even feasible configurations with clear differences in the technology mix unexplored. Previous empirical work showed that configurations with very low or very high shares of a particular technology in the overall capacity mix, located near the `corners', or extremes, of the multi-dimensional decision space, are not found by conventional applications of MGA even when they exist -- that is, when they can be eventually found if the model objective is explicitly set to find them \cite{decarolis_temoa_2016,neumann_near-optimal_2021}. 

Second, to keep the problem computationally tractable, the search for something `as different as possible' is typically applied only to high-level variables of energy system models, such as the total installed capacity for each technology type, even within models of high spatial granularity \cite{neumann_near-optimal_2021}. This leads to the generation of `technologically-distinctive' feasible solutions, which rely relatively more on technologies that were minimally deployed in the least-cost solution, for instance deploying more solar than wind, at the system scale. However, real-world concerns, such as the social acceptance of new infrastructure, primarily call for alternatives in the way capacity is located at the sub-national scale, even for a fixed, agreed-upon mix of technologies at the national scale \cite{lombardi_policy_2020}. For instance, different ways of spatially distributing wind capacity across sub-national regions, for a roughly fixed amount of total wind capacity to be deployed in the system. While it is true that technologically-distinctive feasible solutions are naturally likely to entail different sub-national distributions of capacity as well \cite{neumann_broad_nodate}, this does not necessarily make up for having spatially-distinctive alternatives around a roughly fixed technology mix \cite{lombardi_policy_2020}. On the other hand, an explicit search for such `spatially-distinctive' solutions substantially enlarges the number of potentially interesting alternatives to generate and, thereby, the overall computational burden.

In response to the two limitations above, several recent advancements of MGA with application to energy system modelling have aimed at either spanning the solution space more evenly or explicitly looking for spatially-distinctive solutions; or, in some cases, at achieving both at once. Yet, as mentioned above, the amount of alternatives that may matter for state-of-the-art energy system models of large size is virtually infinite. Even attempts at mapping `all' alternatives \cite{pedersen_modeling_2021} or spanning the solution space as evenly as possible \cite{neumann_broad_nodate} can ultimately provide only a finite sample of solutions based on high-level technological dissimilarity. Furthermore, providing a finite set of options leaves open the question of how many alternatives are enough, i.e. which alternatives are redundant and which ones are not.

Acknowledging this, we set out to empirically investigate the computational trade-offs among different possible approaches to the generation of alternatives within high-resolution energy system models. In particular, we test four different ways of generating alternatives within the SPORES method, an original development of MGA that we have presented in previous work \cite{lombardi_policy_2020}. The use of SPORES does not cover all possible formulations of MGA in the literature; but the high customisability of the SPORES workflow lends itself to tweaking the search towards either spatial or technological dissimilarity explicitly, allowing us to explore the trade-offs between the two. Some of the tweaks of the search strategy that we consider do originate from other recent MGA advancements but are applied here in a spatially-explicit way even when they were originally conceived for application to high-level variables only (see section \ref{methods}). With application to a model of the European power system with 97 nodes and 3-hour temporal resolution, we show that deciding what is redundant and what is not is far from trivial, and that there is a clear trade-off in the computational efficiency of MGA search strategies in highlighting technological versus spatial dissimilarity of feasible solutions. In particular, we show that focussing on technology dissimilarity may fail to identify system configurations that appeal to real-world stakeholders, such as those in which capacity is more spread out at the local scale. Based on our empirical findings, we propose possible solutions for energy modellers to make their MGA approaches as computationally efficient and practically relevant as possible. We openly release the model and code used for the generation of results on Zenodo \cite{lombardi_customised_2022}, to foster transparency, repeatability and further developments \cite{pfenninger_importance_2017}.

The remainder of the paper is organised as follows. Section \ref{methods} presents the latest version of the SPORES workflow, how we tweak its parameters to approximate different search strategies, and the European power system model to which we apply our experiments. Section \ref{results} shows and discusses our findings in terms of the overall discovered decision space for each search strategy, the amount of generated alternatives that are likely relevant for real-world decisions, and differences in the resulting flexibility for spatially moving capacity. We conclude (section \ref{conclusion}) by discussing the implications of our findings for other modellers and policy decision support.

\section{Methods} \label{methods}

The following subsections present the latest version of the SPORES workflow (subsection \ref{methods-workflow}), the different search strategies that we use within the workflow for the analysis of computational trade-offs (\ref{methods-search-strat}), and the specific version of the Euro-Calliope model that we use as a case study (\ref{methods-eurocalliope}).

\subsection{SPORES workflow} \label{methods-workflow}

\begin{figure}[H]
	\centering
	\includegraphics[width=1\textwidth]{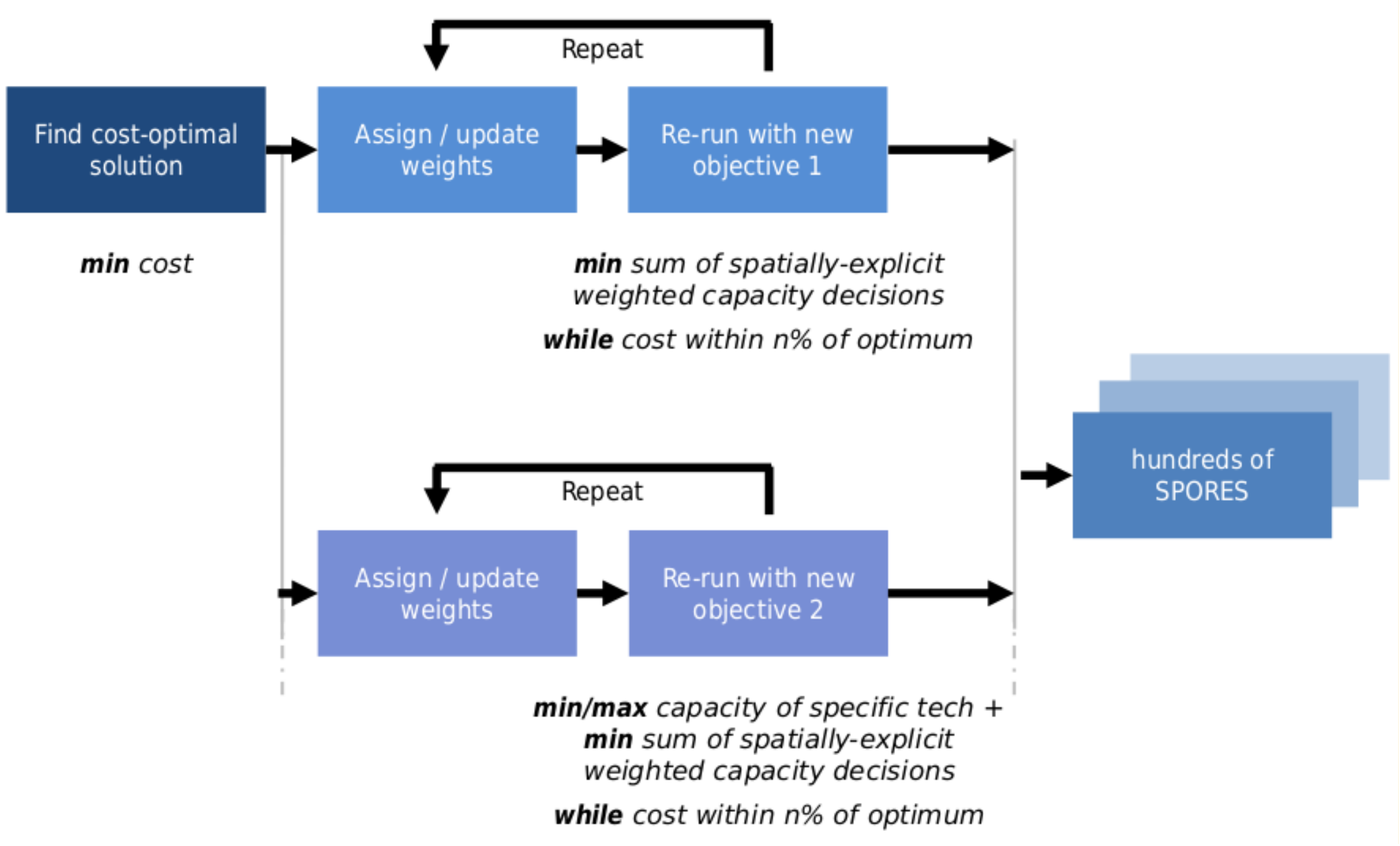}
	\caption{Schematic representation of the SPORES workflow. The feasible, near-optimal solution space is systematically explored through parallel runs. A base run (top of the figure) searches for alternatives adopting a spatially-explicit version of MGA. Weights are assigned not only to system-wide technology-capacity decision variables but rather to each technology-in-location capacity decision variable. In parallel, several other runs (bottom of the figure) explore the decision space from different directions. These runs add to the search for spatially-explicit alternatives a secondary, technology-explicit objective: the system-wide minimisation (and, alternatively, maximisation) of a specific technology. This ensures that configurations with an extremely high or low deployment of specific technologies do not get lost while also ensuring that several spatially-distinctive configurations of technology deployment are provided around those.}
	\label{fig:spores_workflow}
\end{figure}

The SPORES workflow, summarised in Figure \ref{fig:spores_workflow} differs from conventional MGA in two key aspects. 

First, it searches explicitly for spatially-distinctive options. Conventional MGA searches for a new solution as different as possible to the optimal one by assigning weights (penalties) to aggregate capacity variables proportional to their capacity deployment in the economically optimal solution. SPORES, instead, assigns such weights to spatially-explicit capacity variables. For instance, if wind generation is deployed in the cost-optimal solution, SPORES assigns different penalties to wind generation capacity in each location rather than penalising wind generation overall. This means that the search for something different might result in a configuration that has  as much wind generation capacity overall, but distributes it differently in space.

Second, it does so from multiple directions within the feasible, near-optimal solution space. This parallel search from multiple directions arises by anchoring the MGA algorithm to different extremes of the feasible near-optimal space, instead of just using the least-cost solution as the starting point of the search. These additional extremes are identified by explicitly minimising or maximising the system-wide deployment of specific technology-capacity decision variables, a strategy also employed in other recent work \cite{neumann_broad_nodate}. We detail the mathematical steps required to generate alternatives based on such a workflow in the following subsections.

\subsubsection{Identification of the cost-optimal solution} \label{methods-spores-opt}

We identify the cost-optimal solution, or economic optimum of the problem, by minimising the total annualised system cost, as per Equation \ref{eq:cost_opt}.

\begin{equation} \label{eq:cost_opt}
	\begin{array}{*{20}{c}}
		{\min :cost\, = \sum\limits_j {\sum\limits_i {\left( {{c_{fix,ij}}x_{ij}^{cap} + \sum\limits_t {{c_{var,ij}}x_{t,ij}^{prod}} } \right)} } }\\
		{s.t.\;\;\matr{Ax} \le \matr{b}}\\
		{\,\,\,\;\;\matr{x} \ge 0,}
	\end{array}
\end{equation}

where $i$ and $j$ indicate the $i$-th technology type and the $j$-th location of the model; $x_{ij}^{cap}$ is the decision variable pertaining to the installed capacity of the $ij$-th location-technology pair; $x_{t,ij}^{prod}$ is the decision variable related to the power production of the $ij$-th location-technology pair as a function of time; $c_{fix,ij}$, $c_{var,ij}$ are, respectively, the annualised fixed and variable costs per each location-technology pair; $\matr{A}$, $\matr{b}$, are a matrix and a vector of coefficients that build all the physical constraints in combination with the vector $\matr{x}$ of all decision variables.  

\subsubsection{Assignment of weights} \label{methods-spores-weight}

Having identified the mathematical optimum, we assign a strictly positive weight ($w_{ij}^n$) to the generation capacity decision variables (location-technology pairs) that are non-zero. This weight can be assigned using different approaches, which result in different search strategies. In the first published application of SPORES \cite{lombardi_policy_2020}, we assigned weights based on the relative deployment of a technology in a given location compared to the maximum potential for deployment at that location ($x_{ij,max}^{cap}$). The newly-found weight is then summed to the weight obtained in the preceding iteration ($w_{ij}^{n - 1} $), for any iteration other than the first (Equation \ref{eq:weight}). Here, we also test three additional weight assignment methods (see subsection \ref{methods-search-strat}).

\begin{equation} \label{eq:weight}
	w_{ij}^n = w_{ij}^{n - 1} + \frac{{x_{ij}^{cap,n}}}{{x_{ij,max}^{cap}}}
\end{equation}

\subsubsection{Main batch of SPORES} \label{methods-spores-main}

We hence obtain a SPORE by minimising the sum of location-specific weighted capacity decision variables. At the same time, we constrain the total annualised system cost ($cos{t_n}$) to remain in a neighbourhood of the optimal cost ($cos{t_0}$), as per Equation \ref{eq:gen_spores}. 

\begin{equation} \label{eq:gen_spores}
	\begin{array}{*{20}{c}}
		{\min \;Y = \sum\limits_j {\sum\limits_i {{w_{ij}}x_{ij}^{cap}} } }\\
		{s.t.\;\;cos{t_n} \le (1 + s) \cdot cos{t_0}}\\
		{\matr{Ax} \le \matr{b}}\\
		{\matr{x} \ge 0,}
	\end{array}
\end{equation}

where $s$ is the accepted cost relaxation (also known as cost slack). Such a new, MGA-like objective function formally makes the problem an $\epsilon$-constrained multi-objective optimisation: the minimisation of already-deployed location-technology pairs is the explicit objective and cost is the implicit one. This objective function is similar to the one first applied by \citeauthor{decarolis_mga_2011} \cite{decarolis_mga_2011} to energy system models and also used in more recent work \cite{decarolis_temoa_2016}, except for the variables and weights being here spatially-explicit.

\subsubsection{Parallel batches of SPORES with secondary objectives} \label{methods-spores-parallel}

The workflow steps outlined in subsections \ref{methods-spores-opt} to \ref{methods-spores-main} can be as well seen as a spatially-explicit version of the `Hop, Skip and Jump' algorithm. Such an algorithm has been repeatedly shown to struggle with pushing the search for alternatives up to the extreme corners of the multi-dimensional decision space \cite{decarolis_temoa_2016,neumann_near-optimal_2021}. The incorporation of the spatial dimension further exacerbates the problem, as it multiplies the variables at stake and the different feasible system configurations. A spatially-explicit version of the algorithm will focus explicitly on spatial diversity and will further struggle to discover alternatives in which the mix of deployed technologies is radically different. For such reasons, SPORES foresee a further systematic exploration of the decision space from alternative directions in which technology diversity is handled explicitly. They do so by adding to the objective function a second explicit objective: the minimisation (and, alternatively, maximisation) of the capacity of a specific technology.

\begin{equation} \label{eq:extra_spores}
	\begin{array}{*{20}{c}}
		{\min\;(or \max) \;{Y_{2,\bar{i}}} = a \cdot \sum\limits_j{ x_{\overline {i}j }^{cap}} + b \cdot \sum\limits_j {\sum\limits_i {{w_{ij}}x_{ij}^{cap}} } }\\
		{s.t.\;\;cos{t_n} \le (1 + s) \cdot cos{t_0}}\\
		{\matr{Ax} \le \matr{b}}\\
		{\matr{x} \ge 0,}
	\end{array}
\end{equation}

where $x_{\overline {i}j}$ is the capacity decision variable associated with the technology under minimisation (or maximisation), and $a$ and $b$ are the weights associated with the different components of the objective function (which can be customised at need, as discussed in \ref{methods-search-strat}). Other recent work \cite{neumann_near-optimal_2021,neumann_broad_nodate,} has used the minimisation or maximisation of specific technologies as a way to explore the near-optimal solution space, but never in combination with the simultaneous generation of spatially-explicit alternatives, which is a unique feature of SPORES. As anticipated above, the rationale behind the combination of technology-explicit and spatially-explicit objectives is that of maintaining a focus on the discovery of spatially-distinctive options while also making sure that the option space is explored from all the relevant search directions. In other words, ensuring that no technology-distinctive option is left unexplored. If needed, the customisability of the coefficients $a$ and $b$ leaves open the possibility for the modeller to collapse the search into only one of the two objectives, thereby replicating, for instance, the search strategy proposed by \citeauthor{neumann_near-optimal_2021} \cite{neumann_near-optimal_2021}.

This configures the problem as a linearised multi-objective optimisation problem, with two explicit objectives parametrised by the coefficients $a$ and $b$ that add up to the implicit $\epsilon$-constrained cost objective. However, despite the mathematical formulation being multi-objective, the two explicit objectives do not aim to reflect any plausible real-world decision factors, which differentiates the approach from typical applications of multi-objective optimisation \cite{prina_multi-objective_2020}. The primary objective is, in fact, just the search for something different from previous iterations, whilst the secondary objective is only needed here as a technical means to operate the spatially-explicit MGA search from alternate extremes of the decision space. Finally, we do not aim to (and do not) find a Pareto-front of optimal solutions by systematically varying the $a$ and $b$ weights of the different components of the objective function, as typically done in multi-objective optimisation. This is for two reasons. First, the resulting Pareto-optimal solutions would not have any real-world meaning since our fictitious objectives and weights do not have any, either. And second, we explicitly want to look at mathematically sub-optimal solutions beyond a fictitious 'Pareto front' as long as they are within the defined cost slack.

\subsection{Customised search strategies} \label{methods-search-strat}

The SPORES workflow presented throughout subsections \ref{methods-spores-opt} to \ref{methods-spores-parallel} lends itself to customisation. Both the weight-assignment method (Equation \ref{eq:weight}) and the relative strength of the two explicit objectives in the parallel batches of SPORES (Equation \ref{eq:extra_spores}) can be modified at need. Different weight-assignment methods may push the search relatively more on spatial versus technology dissimilarity from previous iterations \cite{lombardi_policy_2020}. Similarly, changing the relative strength of the technology-explicit secondary objective in Equation \ref{eq:extra_spores} may allow the search to depart more or less from the identified extreme of the decision space. In this work, we test four different weight-assignment methods and two levels of relative strength for the technology-explicit secondary objective. In both cases, the aim is to test the outcomes, in terms of redundancy of generated solutions, of pushing the search more on technology versus spatial dissimilarity, or viceversa.

\subsubsection{Alternative weight-assignment methods}

First, we consider the weight-assignment method outlined in Equation \ref{eq:weight}, which we hereafter refer to as the \textit{relative-deployment} method. We designed this method with a focus on spatial dissimilarity in previous work \cite{lombardi_policy_2020}. 

Second, we test the so-called \textit{integer} method originally proposed by \citeauthor{brill_use_1979} \cite{brill_use_1979} and later applied to energy system modelling by \citeauthor{decarolis_mga_2011} \cite{decarolis_mga_2011}. Reported in Equation \ref{eq:weight-integer}, it is the simplest method, but has the potential drawback of many variables ending up with the same weight, hampering the search efficiency. We apply it here in a spatially-explicit way, assigning weights to location-technology pairs rather than to system-wide technology variables only.

\begin{equation} \label{eq:weight-integer}
	w_{ij}^n = w_{ij}^{n - 1} + k_{ij}, \;\; {with}	 \; k_{ij} =
	\begin{cases}	
		100, \; if \; x_{ij} > c\\
		0, \; if \; x_{ij} \le c
	\end{cases}
\end{equation}

where $c$ is a constant threshold defined to avoid that even very marginal deployments of capacity may receive a weight, which would otherwise entail the risk that almost all location-technology pairs receive a non-zero weight. It is also worth noting that the \textit{integer} weight amounts here to 100 based on the internal unit scaling of our model; a different scaling, say 1 or 10, may make more sense for a model with different units.

Third, we consider a \textit{random} method, in which weights have no rationale and are indeed assigned as random integer numbers (Equation \ref{eq:weight-random}). This method approximates the random MGA search proposed by other authors \cite{berntsen_ensuring_2017,sasse_regional_2020}. Unlike such previous work, though, we do not consider further degrees of randomisation in the objective function, for the sake of consistency with the other analysed methods, and we apply the random weights to spatially-explicit decision variables.

\begin{equation} \label{eq:weight-random}
	w_{ij}^n = w_{ij}^{n - 1} + r_{ij}, \;\; {with}	\; r_{ij} = U(0,100)
\end{equation}

where $U(0,100)$ is a random uniform distribution.

Fourth and final, we propose an original \textit{evolving-average} method (Equation \ref{eq:weight-evolving}). The idea of this method is to retain a more explicit memory of past iterations, compared to just having incremental weights. In such a way, one can assign a weight to each location-technology pair based on the distance from the \textit{average} capacity deployed for that pair across all previously found feasible solutions (${\overline{x_{ij}}^{cap,n-1}}$), which is kept up to date -- in other words, it \textit{evolves}.

\begin{equation} \label{eq:weight-evolving}
	\begin{cases}	
			w_{ij}^n = |\ffrac{ {{\overline{x_{ij}}^{cap,n-1}}} - {x_{ij}^{cap,n}} }{{\overline{x_{ij}}^{cap,n-1}}}|\\
		{{\overline{x_{ij}}^{cap,n-1}}} = \ffrac{\sum\limits_{n=1}^{n-1} x_{ij}^{cap,n}}{n-1}
	\end{cases}
\end{equation}

\subsubsection{Relative strength of spatial- and technology-explicit objectives} \label{methods-spores-secondary_obj} 
The parallel batches of SPORES arising from technology-explicit extremes of the decision spaces can be customised by tweaking the parameters $a$ and $b$ of Equation \ref{eq:extra_spores}. By default, both parameters are set to a unitary value, which means they have the same relative strength. Here, we test the outcome of reducing the relative strength ($a$) to 0.1 of the technology-explicit secondary objective. We want to avoid obtaining, for a batch of SPORES generated around the minimisation of a certain technology, only configurations in which such a technology is fully minimised. A reduction of the relative strength of the technology lever may allow us to obtain configurations in which the given technology is only partially minimised, further improving the technology dissimilarity of the generated alternatives. As for the case of the other customised parameters (see subsection \ref{methods-spores-weight}), the absolute numbers we adopt here for the $a$ and $b$ parameters make sense within the unit scaling of our model. Other absolute values might be more appropriate for a different unit scaling. However, since our analysis focuses on modifications of the parameters relative to one another, the outcomes remain generally valid.

\subsection{Energy system model and MGA setup} \label{methods-eurocalliope}

We apply our customisations of the SPORES method to a power system model of Europe comprising 34 countries and 97 nodes, or locations, across those. The model is based on the well-established open-source energy system modelling framework Calliope \cite{pfenninger_calliope_2018} and takes the name of Euro-Calliope. Originally conceived by \citeauthor{trondle_trade-offs_2020} \cite{trondle_trade-offs_2020}, the version we use here is based on a previous work of ours \cite{pickering_diversity_2022} in which we updated the initial (brownfield) electricity grid topology to mirror the most relevant real-world transmission constraints as identified by the e-Highway 2050 project. Besides various hydroelectric technologies whose capacity is assumed to remain constant, the model features seven main technologies for capacity expansion: roof-mounted and open-field solar photovoltaic, on-shore and off-shore wind, bioenergy power plants, and battery and hydrogen-to-power storage. Each of these is used as a basis for two parallel batches of SPORES (Equation \ref{eq:extra_spores}), one in which the deployment of the technology is maximised and one in which it is minimised, at the system level. Transmission technologies are also allowed to expand, and add up to the parallel batches of SPORES. We generate 10 alternatives for each parallel batch, for a total of 160 alternatives. These add up to the 50 alternatives generated within the main batch, leading to an overall 210 SPORES. For all SPORES, we adopt a slack cost of 10\%, in line with previous work \cite{lombardi_policy_2020,pickering_diversity_2022,neumann_near-optimal_2021}. We refer the readers to the Supplementary Methods for further details about the model and how it has been customised for this work.

Although the Euro-Calliope model version that we use in this work allows for the modelling of all energy sectors, we subset the analysis to the power sector alone, without considering the additional electricity demand arising from the likely electrification of sectors such as building heat and transport. In fact, the aim of the present work is to focus on the computational trade-offs between different approaches to the generation of alternative near-optimal solutions. Subsetting the analysis to the power sector alone allows us to test many approaches while keeping the computational effort feasible. It also allows readers to compare our findings with those of other recent MGA studies applied to energy system models of Europe \cite{neumann_near-optimal_2021,pedersen_modeling_2021,neumann_broad_nodate,grochowicz_intersecting_2022}, which are all grounded in a similar model setup. What is more, the analysis of the available decision space for the full decarbonisation of Europe including all energy-consuming sectors has already been undertaken using Euro-Calliope \cite{pickering_diversity_2022} and would therefore be redundant to repeat here.

\section{Results} \label{results}

We present our results for the reference case in which the relative strength of the two SPORES objectives is even (see subsection \ref{methods-spores-secondary_obj}), whilst we provide Supplementary Results for the case of a reduced strength of the secondary, technology-explicit objective. We start by analysing the `shape' of the overall discovered decision space across the four search strategies, discussing which decision space is richer in technologically-distinctive options. Hence, we assess how many of the discovered alternatives, in each case, would match a plausible stakeholder interest, such as the limited concentration of onshore wind farms, discussing the efficiency of discovering many spatially-distinctive options across search strategies. Finally, we look for potentially real-world relevant spatial features of feasible configurations that may not be common to all search strategies and discuss how this may impact the practical usefulness of a search strategy.

\subsection{Overall discovered decision space} \label{results-sporeplots}

\begin{figure}[H]
	\centerline{\includegraphics[width=1\linewidth]{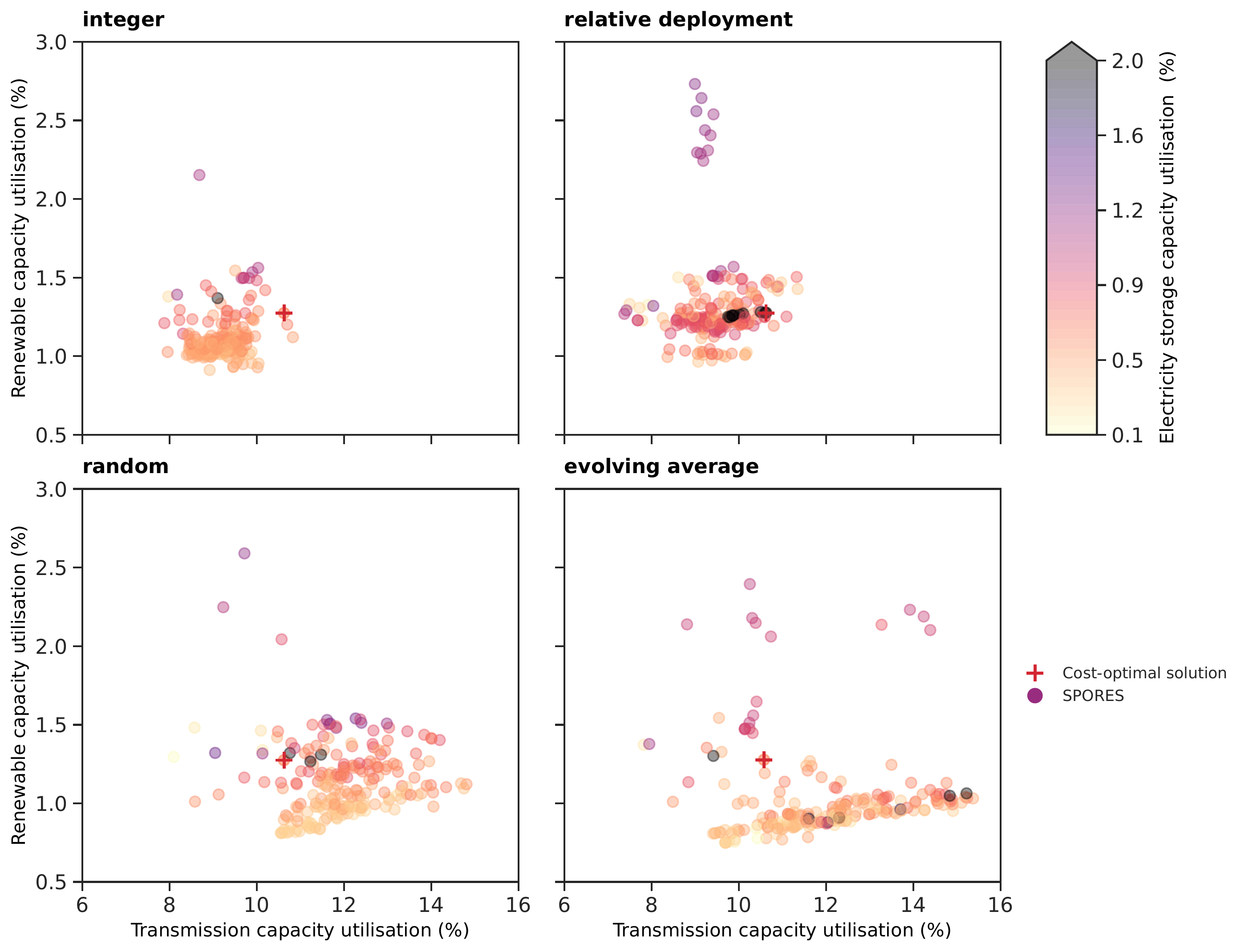}}
	\caption{Generated alternatives in a tri-dimensional space defined by aggregate renewable (Y-axis), transmission (X-axis) and storage (colour coding) capacity deployment compared to maximum deployment potential. Results are provided for all four considered search strategies. The cost-optimal solution is identified in each panel by a red marker. }
	\label{fig:sporeplot}
\end{figure}

The four considered search strategies showcase substantial differences in the overall discovered decision space, as shown in Figure \ref{fig:sporeplot}. These differences can be even more marked if considering only those alternatives generated within the main batch of SPORES (see Figure \ref{fig:sporeplot_main_batch}), but are mitigated when parallel batches with technology-explicit secondary objectives are included, as per the default SPORES workflow presented in subsection \ref{methods-workflow}. This evidence reinforces the importance of systematically exploring the decision space from multiple directions, in line with recent work \cite{lombardi_policy_2020,neumann_broad_nodate}. 

Regarding trade-offs between spatial versus technology dissimilarity, the \textit{integer} and \textit{relative-deployment} strategies tend to produce less-sparse solutions, with many alternatives almost overlapping in the tri-dimensional space that considers total renewables, transmission and storage capacity deployed. Conversely, the \textit{random} and the \textit{evolving-average} strategies seem to produce more sparsity and push the alternatives away from each other more, particularly along the transmission-expansion axis, despite both still ending up with a non-negligible number of overlapping solutions. This means that the \textit{random} and evolving average strategies are more efficient at discovering markedly distinctive options from a technology perspective, whilst the \textit{relative-deployment} and \textit{integer} methods focus more on dissimilarity of spatial deployment around fewer distinctive overall technology mixes. In fact, the alternatives overlapping in the tri-dimensional space outlined in Figure \ref{fig:sporeplot} must not be mistaken for identical, and hence redundant, solutions. On the contrary, they are solutions that, albeit similar in terms of the total deployed capacity of the different technology options, are likely radically different in terms of their spatial configuration of technology deployment. Figure \ref{fig:stripviolin_1}, which expands the cross-search-strategy comparison by looking at the deployment of further disaggregated technologies, confirms the same trend: the \textit{relative-deployment} and \textit{integer} search strategies showcase a higher degree of overlapping solutions, which is particularly apparent for wind generation and transmission capacities, but generally valid for all technologies.

Differences across methods do not change qualitatively for the case in which the secondary objectives aimed at minimising or maximising specific technologies are assigned a relatively weaker weight in the objective function (Figure \ref{fig:sporeplot_01}, complemented by Figure \ref{fig:stripviolin_01} for technology-disaggregated results). However, as expected, the number of alternatives located at the extremes of the decision space -- i.e., those focussing on generating different spatial configurations around an explicit high-level technology feature, say the minimal deployment of bioenergy -- decreases slightly overall. To further investigate the trade-off between technology and spatial dissimilarity, we move on to looking at metrics related to spatial aspects and real-world concerns that play out on a more local scale.

\subsection{Alternatives that match plausible stakeholder interests} \label{results-useful-spores}

\begin{figure}[H]
	\centering
	\includegraphics[width=1\linewidth]{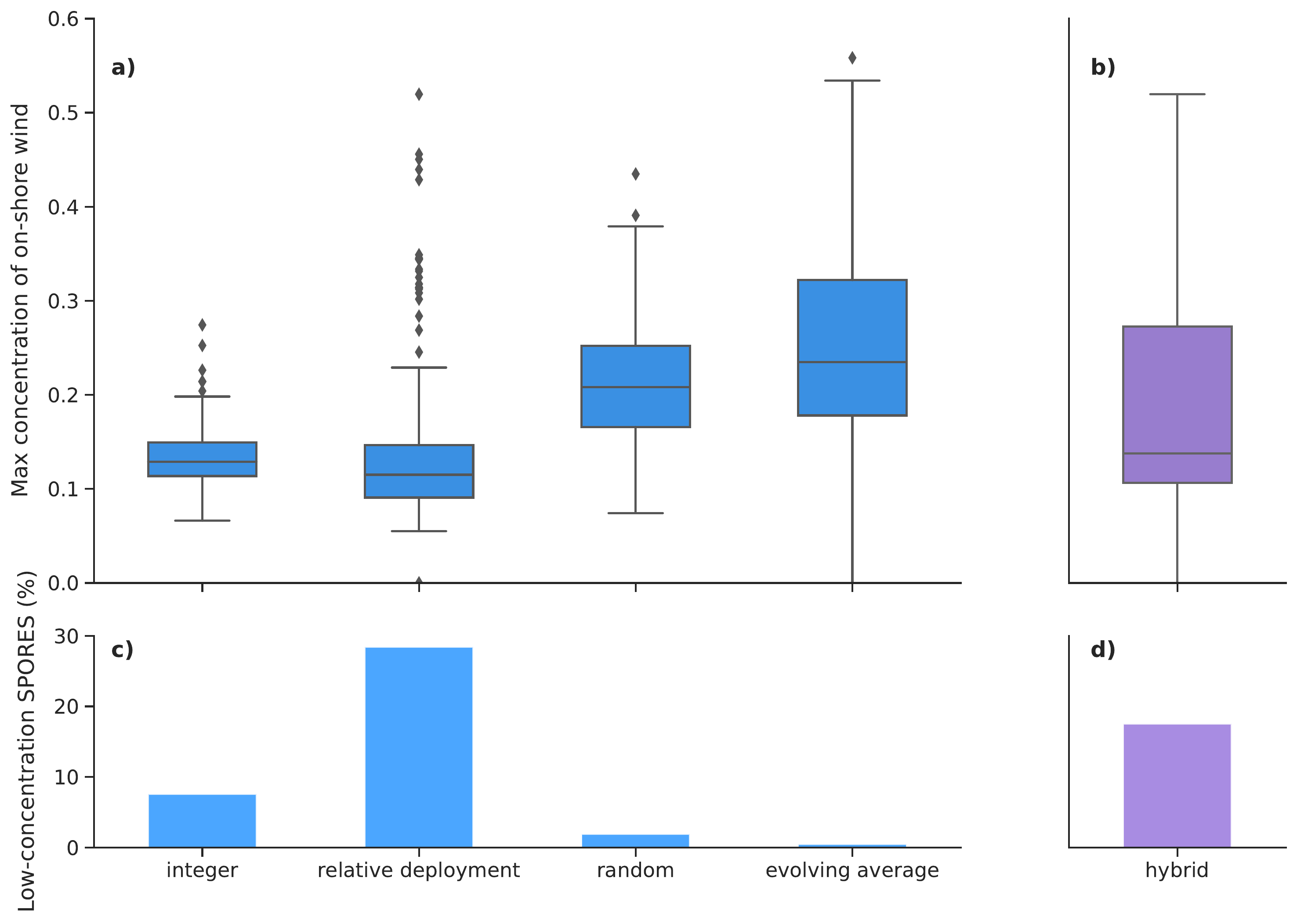}
	\caption{ The top of the Figure shows the data distribution (boxplots) for the maximum concentration of deployed on-shore wind capacity in a single region (across all 97 model locations) compared to total deployed on-shore wind capacity. The boxplots are shown for the four tested weight-assignment methods (panel \textbf{a}) and for a \textit{hybrid} case that combines ex-post the data from the \textit{relative-deployment} and \textit{evolving-average} methods (panel \textbf{b}); see subsection \ref{results-useful-spores} for further details about this \textit{hybrid} case. The bottom of the Figure shows the fraction of generated alternatives with a concentration of on-shore wind capacity at least 20\% lower than in the cost-optimal case. Results are again shown for the four tested weight-assignment methods (panel \textbf{c}) and for the \textit{hybrid} case (\textbf{d}). }
	\label{fig:boxplots}
\end{figure}

One of the most common real-world concerns when deploying new infrastructure for the energy transition is the social acceptability of on-shore wind farms \cite{price_implications_2020}. It is thus helpful to analyse how many of the generated feasible energy system configurations ensure a reduction in the maximum concentration of on-shore wind farms in any single region relative to the total deployed on-shore wind capacity. Such a metric is of the possible proxies for the richness of spatial deployment options, allowing us to investigate further whether those search strategies that performed less well in terms of technology dissimilarity actually provide richer insights regarding the decision flexibility for moving capacity spatially. If a search strategy produces relatively more alternatives with a high concentration of wind farms, that means it focuses primarily on reducing wind capacity overall or on moving highly-concentrated on-shore wind hubs elsewhere without spreading capacity out. Figure \ref{fig:boxplots}a shows the distribution of the maximum on-shore wind concentration across all the SPORES generated by each search strategy. As hinted at by the results in the previous subsection, search strategies that are less efficient at providing technologically-distinctive solutions (such as the \textit{integer} and \textit{relative-deployment} ones) are those with lower median values for on-shore wind farm concentration in single regions. In other words, they generate many solutions in which wind capacity is sited differently across sub-national regions, eventually leading also to solutions in which capacity is more spread out, which is likely of particular importance when providing practical alternatives to real-world acceptability concerns. 

Instead, the search strategies with the wider distribution of values for on-shore wind capacity concentration are the \textit{relative-deployment} and the \textit{evolving-average} ones. However, the median value for the \textit{evolving-average} distribution lies much higher than that of the \textit{relative-deployment} one. This suggests that the \textit{evolving-average} method is highly efficient at spanning the full spectrum of wind concentration options, but -- given the finite number of alternatives -- does so at the expense of generating further spatial dissimilarity around the found configurations. In other words, there is a trade-off between using the limited computational power and time to efficiently span all the most radically different options from a high-level perspective versus generating many feasible spatial configurations in the `technology-neighbourhood' of fewer feasible system configurations.

Another way to quantify this trade-off is to count, for each search strategy, how many of the generated alternatives match the plausible stakeholder interest of limiting on-shore wind capacity concentration. For instance, filtering out alternatives in which on-shore wind capacity is at least 20\% less concentrated than in the least-cost case, as done in previous work \cite{lombardi_policy_2020}. The outcomes of such filtering (Figure \ref{fig:boxplots}c) confirm that the \textit{integer} and, primarily, \textit{relative-deployment} search strategies outperform the others from a spatial-dissimilarity perspective. This is in line with our expectations since the \textit{relative-deployment} weight-assignment method was designed precisely to focus more on spatial dissimilarity (see subsection \ref{methods-spores-weight}). The results of the \textit{random} and, particularly, \textit{evolving-average} approaches further strengthen the finding that the more a method is efficient at spanning high-level system configurations, the less it is efficient at producing spatially-distinctive, non-concentrated system configurations. 

Based on this apparent trade-off, combining the benefits and limitations of different methods into a \textit{hybrid} search may be helpful. For instance, the main batch of SPORES in which the focus is on spatial dissimilarity may be generated via the \textit{evolving-average} method, which will make sure to span an even range of technology options. The parallel batches, in which technology dissimilarity is handled explicitly, may instead use a \textit{relative-deployment} method to ensure that configurations are also spatially distinctive. The outcomes of such a \textit{hybrid} method are shown in Figure \ref{fig:boxplots}b,d, resulting in a good compromise in terms of stakeholder-appealing, low-wind-concentration solutions. 

The results do not change substantially for a weaker anchoring of the search to the extremes of the decision space, except for the \textit{relative-deployment} search strategy that experiences a reduction in the total amount of alternatives with a low concentration of wind farms (Figure \ref{fig:boxplots_01}). This peculiarity is coherent with our finding so far that the \textit{relative-deployment} method is the one that most efficiently explores the flexibility for moving capacity spatially. The \textit{relative-deployment} search strategy takes advantage of his high spatial focus to generate many spatially-distinctive alternatives around extreme technology features of the decision space within parallel batches of SPORES (see sub-section \ref{methods-spores-parallel}). When the anchoring to extreme technology features becomes weaker, the search departs too quickly from the given extreme technology feature for the \textit{relative-deployment} method to eventually generate low-concentration alternatives around it. Other weight-assignment methods are less affected by the same phenomenon because they more naturally tend to depart from the extreme technology feature regardless of the strength of the anchoring. Such a worsening of the performance affects, in turn, also the \textit{hybrid} method. Fine-tuning the relative strength of the two objectives in Equation \ref{eq:extra_spores} to the chosen search strategy appears thus essential to maximise the performance.

\subsection{Unique spatial features} \label{results-maps}

We have discussed the concentration of on-shore wind capacity as an example of a plausible criterion for stakeholder discussion and observed a difference across search strategies in the number of alternatives which limit it. This difference can have very concrete implications. In line with previous work \cite{lombardi_policy_2020,pickering_diversity_2022}, let us assume that stakeholders may be interested in filtering out the decision space based on multiple criteria at once. For instance, minimising the deployment of bioenergy power plants while simultaneously reducing on-shore wind capacity concentration (with the same threshold as defined in subsection \ref{results-useful-spores}). In this case, the decision spaces discovered by the four search strategies would provide very different outcomes. 

\begin{figure}
  \vspace*{-2cm}
  \noindent\makebox[\textwidth]{\includegraphics[width=1.1\textwidth]{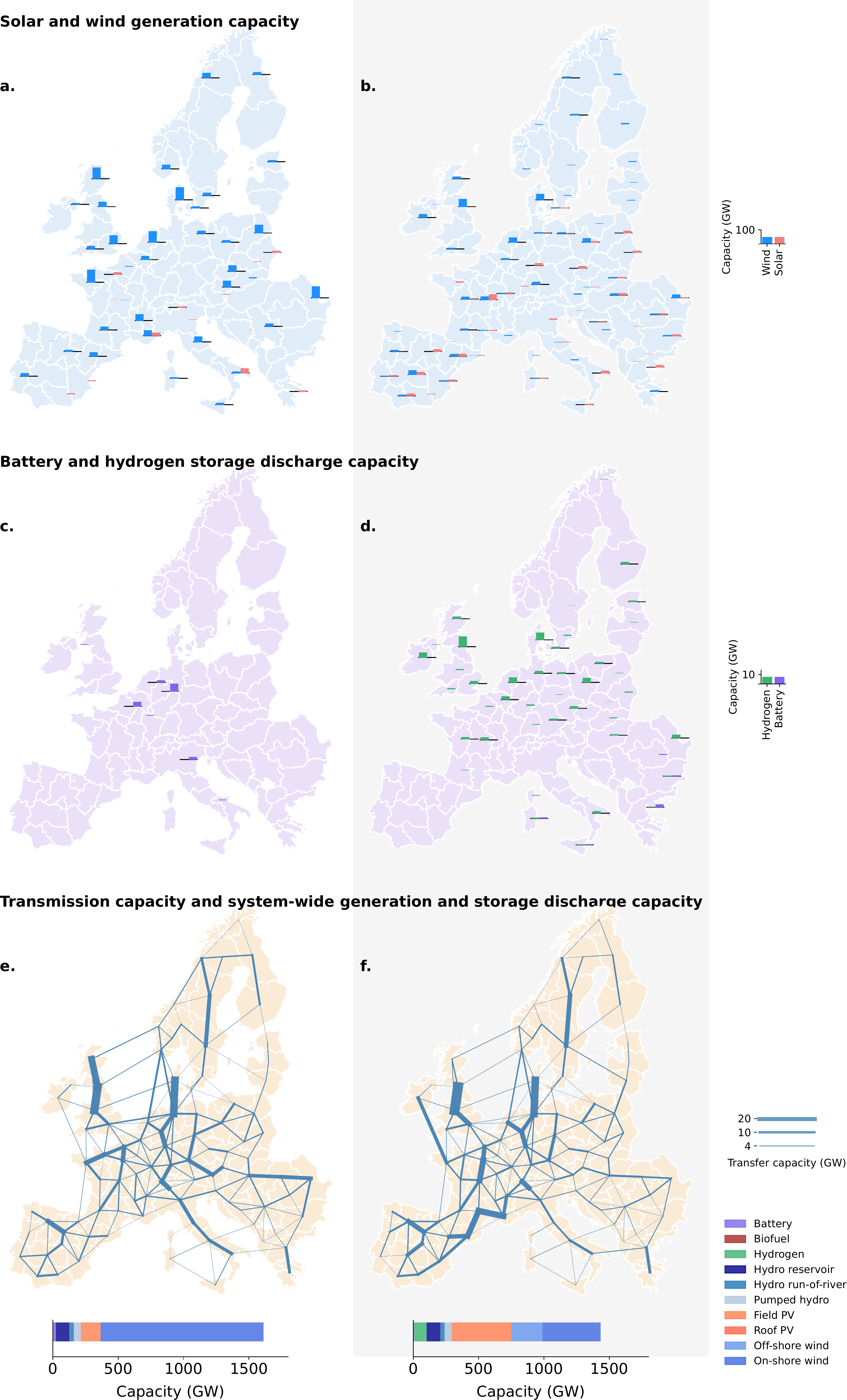}}
  \caption{SPORES with the lowest concentration of on-shore wind capacity and simultaneous minimisation of bioenergy capacity in the \textit{random} (left) and \textit{relative-deployment} (right, grey background) solution spaces. Panels \textbf{a--b} show the deployed capacity of solar (rooftop plus open-field PV) and wind (on-shore plus off-shore) in each of the 97 model locations where the locally-deployed capacity of at least 10 GW. Panels \textbf{c--d} show the spatial deployment of battery and hydrogen storage (discharge) capacity where the locally-deployed capacity is of at least 1.5 GW. Panels \textbf{e--f} show the capacity of transmission lines connecting each location, including both existing and expanded line capacity. Panels \textbf{g--h} show the system-wide capacity deployed for all considered generation and storage technologies.}
  \label{fig:maps}
\end{figure}

The feasible solutions found by the \textit{integer} and \textit{evolving-average} strategies do not feature any option that satisfies the above criteria at once. Minimising bioenergy appears to be only possible if accepting large, concentrated on-shore wind power hubs in key regions. Conversely, both the \textit{relative-deployment} and \textit{random} search strategies include system configurations that satisfy the chosen criteria. Figure \ref{fig:maps} shows, in particular, the feasible system configurations with the lowest concentration of on-shore wind farms that simultaneously allow the reliance on bioenergy to be completely avoided for both the \textit{random} and the \textit{relative-deployment} solution spaces. Albeit both homogeneous in their wind capacity deployment, the configuration found via \textit{random} search (Figure \ref{fig:maps}a,c,e,g) has about three times higher on-shore wind deployment overall. This results in several regions still becoming large wind power hubs, although few to no regions carry an unfair share of total capacity deployment. Instead, the solution discovered by the \textit{relative-deployment} method (Figure \ref{fig:maps}b,d,f,h) is substantially more balanced in terms of technology mix, with off-shore wind and solar capacities being deployed alongside a spatially-homogeneous deployment of on-shore wind capacity. Substantial hydrogen storage and further grid reinforcements towards the Iberian peninsula support the balancing of this configuration. In other words, in this arbitrary yet plausible example, a search strategy that seemed not to be particularly efficient at spanning the solution space if considered from a high-level technology perspective, ends up being the only one capable of identifying a solution that meets certain stakeholder preferences when looked at from the perspective of spatial infrastructure deployment.

The situation changes when there is less of a push towards searching the extremes of the decision space. In this case, no weight-assignment method can discover a configuration that simultaneously avoids bioenergy and limits the concentration of on-shore wind capacity. This is consistent with our expectations. As anticipated in subsection \ref{methods-spores-secondary_obj} and also observed in subsection \ref{results-useful-spores}, the resulting decision space features fewer solutions in which a given technology -- say, bioenergy -- is fully minimised, and more in which its deployment is only partially reduced. Accordingly, there are no more of those apparently redundant alternatives in which a key part of the technology mix remains the same (i.e., no bioenergy) while capacity is deployed differently from a spatial perspective.

As discussed above, we have selected the example of a desire to simultaneously limit the concentration of wind generation capacity and the reliance on bioenergy based on previous work that identified those as plausible stakeholder interests. However, this is just one illustrative example out of many possible ones, motivated by previous work that identified these criteria as particularly appealing to real-world stakeholders \cite{price_implications_2020,lombardi_policy_2020}. A similar case could be made for other plausible combinations of stakeholder interests, say, the degree and topology of expansion of transmission lines and the type and amount of storage technologies deployed. 

\section{Discussion and conclusion} \label{conclusion}

Our empirical comparison of different mathematical strategies for the generation of near-optimal alternatives within large-scale, high-resolution power system models aimed at investigating how to make the best use of limited computational power. We wanted to assess which search strategies may generate redundant alternatives and which ones may not. Our findings allow us to conclude the following.

First, the boundaries of the overall decision space do not change much with the selected search strategy, provided that the extremes of such a decision space are systematically explored. This corroborates the findings of other recent studies \cite{neumann_broad_nodate,pickering_diversity_2022}. 

Second, for an arbitrarily fixed number of generated alternatives (i.e., for a given accepted computational effort), there is a clear trade-off between making use of the computational power to discover alternatives in terms of the high-level technology mix; and using it to generate different spatial configurations of technology deployment in the technology-neighbourhood of fewer high-level technology mixes. The research question should hence guide the choice of the search strategy. If the research focus requires an as-homogeneous-as-possible exploration of the high-level technology mixes, the MGA search can be targeted to such an aim. For instance, relying on the \textit{evolving-average} search strategy proposed in this work or using a brute-force optimisation that targets technology mixes \cite{neumann_broad_nodate}. Yet, such a high-level technology focus is likely -- more so, the more limited the computational power available -- to miss out on most of the stakeholder appealing, spatially-distinctive system configurations around the found technology mixes. Accordingly, whenever research aims at providing alternatives to support real-world decisions, spatial dissimilarity should be at the core of the search strategy. This corroborates our observation in section \ref{intro}, based on the literature, that no single method can capture everything, simply because the set of potentially relevant options is infinite.

Third, as a consequence of the above findings, and particularly when MGA is used to support practical decisions, the computational workflow should foresee iterations with the relevant stakeholders. In such a way, the initial, inherently inexhaustive decision space can be refined based on stakeholder feedback, redirecting the available computational power specifically towards their interests and needs. In this framework, it might be ideal to set up the search in a way that initially compromises between spatial and technology dissimilarity, providing a practically helpful overview of the options. The \textit{hybrid} workflow proposed in subsection \ref{results-useful-spores} might be the most suited for such an initial exploration. Another option, albeit more computationally burdensome, could be to integrate the abovementioned brute-force methods for the homogeneous exploration of technology mixes with the search for a few spatially-distinctive solutions around each found technology mix. Regardless of the chosen method, stakeholders may then indicate themselves which additional technology combinations they would like to investigate that are not initially available, or for which of the existing technology features they would like to see more options to locate capacity spatially. The conceptualisation of a coherent method for creating such a ‘human-computer feedback loop’ is one further development of our method that we are investigating in the context of an ongoing project, SEEDS\footnote[1]{\url{https://seeds-project.org}}.

Overall, our results demonstrate that using modelling analyses to outline viable deployment strategies is, in practice, a challenging task. Previous studies have exposed the pitfalls of the common practice of relying on single, cost-optimal results, and have emphasised the potential of MGA to mitigate the provision of misleading insights \cite{decarolis_temoa_2016,berntsen_ensuring_2017,neumann_near-optimal_2021,lombardi_policy_2020}. Our study does confirm that MGA has the potential to provide more meaningful and robust samples of the practically feasible decision space, particularly when carried out from multiple search directions, as done in the most recent literature \cite{lombardi_policy_2020,pickering_diversity_2022,neumann_broad_nodate}. Nonetheless, our results also warn that MGA is not a panacea per se and needs to go along with careful modeller judgement. For state-of-the-art energy system models of large size, even around 200 samples of near-optimal alternatives – let alone the standard provision of a single, optimal solution – may not be enough to robustly assess whether a particular system configuration is not feasible near the economic optimum. Our arbitrary yet highly plausible example of the configuration that minimises bioenergy while limiting on-shore wind capacity concentration is one such case in point. For instance, let us assume we had provided results based on only one of the search strategies tested in this work, say the \textit{integer} one. We might have been tempted to conclude that minimising the use of bioenergy in a fully carbon-neutral European power system is only possible if accepting large, concentrated on-shore wind power hubs. Yet, we have seen that other search strategies lead to a different conclusion. Besides increasing the sample size and search directions as we do in a recent study \cite{pickering_diversity_2022} and adopting more balanced search strategies like the \textit{hybrid} one proposed in this work, a computationally inexpensive solution to improve the robustness of modelling results might be the adoption of ad-hoc MGA sensitivity analyses targeted to specific policy-relevant claims. In the example above, we could have customised the MGA search to explicitly look for a few solutions with minimal bioenergy deployment and constrained concentration of on-shore wind capacity, finding out that those do, in fact, exist. Such a simple MGA-based counterfactual experiment may help corroborate modelling results both in conventional cost-optimisation studies and in cases in which high-performance computing facilities are not available and carrying out a very large MGA analysis is impossible. 

A less immediate option to improve the computational viability of MGA analyses for complex energy system design problems could be exploring radically different algorithms, such as heuristic ones. These typically iterate by moving from one sub-optimal solution to an improved one, gradually converging to the optimum. It may be thus sufficient to retain those sub-optimal solutions in memory to obtain both the optimum and the near-optimal alternatives with a single model run. So far, heuristics like particle swarm, genetic algorithms and others have been primarily employed in the context of multi-objective optimisation to find sets of Pareto-optimal energy system design options between two explicit real-world objectives \cite{prina_classification_2020}. We have here argued that this type of optimisation faces limits when applied to complex energy transition questions because of the countless stakeholders and unmodelled objectives involved. However, the same underlying algorithms may be repurposed explicitly towards the generation of alternatives without explicit real-world objectives, as is the goal in MGA. This may allow overcoming the limitations of conventional multi-objective optimisation while retaining the advantage of heuristic algorithms. In addition, depending on the chosen approach, heuristic algorithms may allow for non-linear problem formulations, which cannot be achieved with deterministic optimisation for problems of very large size, such as those we discuss here. Future research should further explore the applicability of these methods to energy planning problems.

Regarding multi-objective optimisation, we have argued that MGA is designed to encompass a broad range of unmodelled objectives, thereby overcoming the limits associated with the finite objectives required by multi-objective methods. Nonetheless, it is worth noting how this is only valid when the applied cost relaxation is large enough to encompass Pareto-optimal solutions located far away from the economic objective. Furthermore, the broader the cost-relaxation space, the higher the number of alternatives required to explore it in a balanced way. For limited computational power and a narrow cost relaxation, multi-objective optimisation might still lead to solutions that lie outside the samples obtained via MGA, for instance, due to being substantially more costly than the selected cost relaxation allows. The research question should guide the choice between MGA and multi-objective optimisation.

Finally, our findings arise from the case study of a highly-resolved European power system, which we select as an ideal example of a 'state-of-the-art, large-scale energy system model' that requires advancements in how MGA is applied. However, the conclusions we draw based on such findings are not case-specific. On the contrary, they are valid for any other energy system model of similar (or higher) technical and spatial detail -- for instance, the model of another continental power system; or the model of a country's energy system at a highly granular sub-national resolution. Any such model would face a similar trade-off between spatial and technical dimensions when using limited computational power for generating alternatives. 

\section*{Acknowledgements}
This work was funded by the SEEDS project. The SEEDS project is supported by the CHIST-ERA grant CHIST-ERA-19-CES-004, the Swiss National Science Foundation grant number 195537, the Fundação para a Ciência e Tecnologia (FCT) grant number CHIST-ERA/0005/2019, the Plan Estatal de Investigación Científica y Técnica y de Innovación 2017-2020, Programación Conjunta Internacional 2020 through the Agencia Estatal de Investigación, and the Estonian Research Council grant number 4-8/20/26. Model runs were performed on the ETH Euler cluster.

\microtypesetup{protrusion=false}
\printbibliography

@article{susser_model_policymaking,
title = {Model-based policymaking or policy-based modelling? How energy models and energy policy interact},
journal = {Energy Research \& Social Science},
volume = {75},
pages = {101984},
year = {2021},
issn = {2214-6296},
doi = {https://doi.org/10.1016/j.erss.2021.101984},
url = {https://www.sciencedirect.com/science/article/pii/S2214629621000773},
author = {Diana S{\"u}sser and Andrzej Ceglarz and Hannes Gaschnig and Vassilis Stavrakas and Alexandros Flamos and George Giannakidis and Johan Lilliestam},
}

@article{susser_user_needs_complexity,
title = {Better suited or just more complex? On the fit between user needs and modeller-driven improvements of energy system models},
journal = {Energy},
volume = {239},
pages = {121909},
year = {2022},
issn = {0360-5442},
doi = {https://doi.org/10.1016/j.energy.2021.121909},
url = {https://www.sciencedirect.com/science/article/pii/S0360544221021575},
author = {Diana S{\"u}sser and Hannes Gaschnig and Andrzej Ceglarz and Vassilis Stavrakas and Alexandros Flamos and Johan Lilliestam},
}

@article{yue_review_2018,
	title = {A review of approaches to uncertainty assessment in energy system optimization models},
	volume = {21},
	issn = {2211-467X},
	url = {https://www.sciencedirect.com/science/article/pii/S2211467X18300543},
	doi = {10.1016/j.esr.2018.06.003},
	language = {en},
	urldate = {2021-11-11},
	journal = {Energy Strategy Reviews},
	author = {Yue, Xiufeng and Pye, Steve and DeCarolis, Joseph and Li, Francis G. N. and Rogan, Fionn and Gallach\'{o}ir, Brian \'{O}.},
	month = aug,
	year = {2018},
	pages = {204--217},
}

@article{decarolis_formalizing_2017,
	title = {Formalizing best practice for energy system optimization modelling},
	volume = {194},
	issn = {0306-2619},
	url = {https://www.sciencedirect.com/science/article/pii/S0306261917302192},
	doi = {https://doi.org/10.1016/j.apenergy.2017.03.001},
	journal = {Applied Energy},
	author = {DeCarolis, Joseph and Daly, Hannah and Dodds, Paul and Keppo, Ilkka and Li, Francis and McDowall, Will and Pye, Steve and Strachan, Neil and Trutnevyte, Evelina and Usher, Will and Winning, Matthew and Yeh, Sonia and Zeyringer, Marianne},
	year = {2017},
	pages = {184--198},
}

@article{price_implications_2020,
	title = {The implications of landscape visual impact on future highly renewable power systems: a case study for {Great} {Britain}},
	issn = {1558-0679},
	shorttitle = {The implications of landscape visual impact on future highly renewable power systems},
	doi = {10.1109/TPWRS.2020.2992061},
	journal = {IEEE Transactions on Power Systems},
	author = {Price, James and Mainzer, Kai and Petrovic, Stefan and Zeyringer, Marianne and McKenna, Russell},
	year = {2020},
	note = {Conference Name: IEEE Transactions on Power Systems},
	pages = {1--1},
}

@article{ellenbeck_how_2019,
	title = {How modelers construct energy costs: {Discursive} elements in {Energy} {System} and {Integrated} {Assessment} {Models}},
	volume = {47},
	issn = {2214-6296},
	shorttitle = {How modelers construct energy costs},
	url = {https://www.sciencedirect.com/science/article/pii/S2214629618306546},
	doi = {10.1016/j.erss.2018.08.021},
	language = {en},
	urldate = {2021-11-11},
	journal = {Energy Research \& Social Science},
	author = {Ellenbeck, Saskia and Lilliestam, Johan},
	month = jan,
	year = {2019},
	pages = {69--77},
}

@article{brown_response_2018,
	title = {Response to ‘{Burden} of proof: {A} comprehensive review of the feasibility of 100\% renewable-electricity systems’},
	volume = {92},
	issn = {1364-0321},
	shorttitle = {Response to ‘{Burden} of proof},
	url = {https://www.sciencedirect.com/science/article/pii/S1364032118303307},
	doi = {10.1016/j.rser.2018.04.113},
	language = {en},
	urldate = {2021-11-11},
	journal = {Renewable and Sustainable Energy Reviews},
	author = {Brown, T. W. and Bischof-Niemz, T. and Blok, K. and Breyer, C. and Lund, H. and Mathiesen, B. V.},
	month = sep,
	year = {2018},
	pages = {834--847},
}

@article{deng_review_2020,
	title = {Power system planning with increasing variable renewable energy: {A} review of optimization models},
	volume = {246},
	issn = {0959-6526},
	shorttitle = {Power system planning with increasing variable renewable energy},
	url = {https://www.sciencedirect.com/science/article/pii/S0959652619338326},
	doi = {10.1016/j.jclepro.2019.118962},
	language = {en},
	urldate = {2021-11-16},
	journal = {Journal of Cleaner Production},
	author = {Deng, Xu and Lv, Tao},
	month = feb,
	year = {2020},
	pages = {118962},
}

@article{prina_multi-objective_2020,
	title = {Multi-objective investment optimization for energy system models in high temporal and spatial resolution},
	volume = {264},
	issn = {0306-2619},
	url = {https://www.sciencedirect.com/science/article/pii/S0306261920302403},
	doi = {10.1016/j.apenergy.2020.114728},
	language = {en},
	urldate = {2021-11-16},
	journal = {Applied Energy},
	author = {Prina, Matteo Giacomo and Casalicchio, Valeria and Kaldemeyer, Cord and Manzolini, Giampaolo and Moser, David and Wanitschke, Alexander and Sparber, Wolfram},
	month = apr,
	year = {2020},
	pages = {114728},
}

@article{brill_use_1979,
	title = {The {Use} of {Optimization} {Models} in {Public}-{Sector} {Planning}},
	volume = {25},
	issn = {0025-1909},
	url = {https://pubsonline.informs.org/doi/abs/10.1287/mnsc.25.5.413},
	doi = {10.1287/mnsc.25.5.413},
	number = {5},
	urldate = {2021-11-16},
	journal = {Management Science},
	author = {Brill, E. Downey},
	month = may,
	year = {1979},
	note = {Publisher: INFORMS},
	pages = {413--422},
}

@article{decarolis_mga_2011,
	title = {Using modeling to generate alternatives ({MGA}) to expand our thinking on energy futures},
	volume = {33},
	issn = {0140-9883},
	url = {https://www.sciencedirect.com/science/article/pii/S0140988310000721},
	doi = {10.1016/j.eneco.2010.05.002},
	language = {en},
	number = {2},
	urldate = {2021-11-16},
	journal = {Energy Economics},
	author = {DeCarolis, Joseph F.},
	month = mar,
	year = {2011},
	pages = {145--152},
}

@article{decarolis_temoa_2016,
	title = {Modelling to generate alternatives with an energy system optimization model},
	volume = {79},
	issn = {1364-8152},
	url = {https://www.sciencedirect.com/science/article/pii/S1364815215301080},
	doi = {10.1016/j.envsoft.2015.11.019},
	language = {en},
	urldate = {2021-11-16},
	journal = {Environmental Modelling \& Software},
	author = {DeCarolis, J. F. and Babaee, S. and Li, B. and Kanungo, S.},
	month = may,
	year = {2016},
	pages = {300--310},
}

@article{neumann_near-optimal_2021,
	title = {The near-optimal feasible space of a renewable power system model},
	volume = {190},
	issn = {0378-7796},
	url = {https://www.sciencedirect.com/science/article/pii/S0378779620304934},
	doi = {10.1016/j.epsr.2020.106690},
	language = {en},
	urldate = {2021-11-16},
	journal = {Electric Power Systems Research},
	author = {Neumann, Fabian and Brown, Tom},
	month = jan,
	year = {2021},
	pages = {106690},
}

@article{lombardi_policy_2020,
	title = {Policy {Decision} {Support} for {Renewables} {Deployment} through {Spatially} {Explicit} {Practically} {Optimal} {Alternatives}},
	volume = {4},
	issn = {2542-4785, 2542-4351},
	url = {https://www.cell.com/joule/abstract/S2542-4351(20)30348-2},
	doi = {10.1016/j.joule.2020.08.002},
	language = {English},
	number = {10},
	urldate = {2021-11-16},
	journal = {Joule},
	author = {Lombardi, Francesco and Pickering, Bryn and Colombo, Emanuela and Pfenninger, Stefan},
	month = oct,
	year = {2020},
	note = {Publisher: Elsevier},
	pages = {2185--2207},
}

@article{li_investment_2017,
	title = {Investment appraisal of cost-optimal and near-optimal pathways for the {UK} electricity sector transition to 2050},
	volume = {189},
	issn = {03062619},
	url = {https://linkinghub.elsevier.com/retrieve/pii/S0306261916318104},
	doi = {10.1016/j.apenergy.2016.12.047},
	language = {en},
	urldate = {2021-11-16},
	journal = {Applied Energy},
	author = {Li, Francis G.N. and Trutnevyte, Evelina},
	month = mar,
	year = {2017},
	pages = {89--109},
}

@article{berntsen_ensuring_2017,
	title = {Ensuring diversity of national energy scenarios: {Bottom}-up energy system model with {Modeling} to {Generate} {Alternatives}},
	volume = {126},
	issn = {03605442},
	shorttitle = {Ensuring diversity of national energy scenarios},
	url = {https://linkinghub.elsevier.com/retrieve/pii/S0360544217304097},
	doi = {10.1016/j.energy.2017.03.043},
	language = {en},
	urldate = {2021-11-16},
	journal = {Energy},
	author = {Berntsen, Philip B. and Trutnevyte, Evelina},
	month = may,
	year = {2017},
	pages = {886--898},
}

@article{price_modelling_2017,
	title = {Modelling to generate alternatives: {A} technique to explore uncertainty in energy-environment-economy models},
	volume = {195},
	issn = {0306-2619},
	shorttitle = {Modelling to generate alternatives},
	url = {https://www.sciencedirect.com/science/article/pii/S0306261917302957},
	doi = {10.1016/j.apenergy.2017.03.065},
	anguage = {en},
	urldate = {2021-11-16},
	journal = {Applied Energy},
	author = {Price, James and Keppo, Ilkka},
	month = jun,
	year = {2017},
	pages = {356--369},
}

@article{neumann_broad_nodate,
	title = {Broad {Ranges} of {Investment} {Con} gurations for {Renewable} {Power} {Systems}, {Robust} to {Cost} {Uncertainty} and {Near}-{Optimality}},
	language = {en},
	author = {Neumann, Fabian and Brown, Tom},
        journal = {arXiv:2111.14443 [physics.soc-ph]},
}

@article{pedersen_modeling_2021,
	title = {Modeling all alternative solutions for highly renewable energy systems},
	volume = {234},
	issn = {0360-5442},
	url = {https://www.sciencedirect.com/science/article/pii/S0360544221015425},
	doi = {10.1016/j.energy.2021.121294},
	language = {en},
	urldate = {2021-11-16},
	journal = {Energy},
	author = {Pedersen, Tim T. and Victoria, Marta and Rasmussen, Morten G. and Andresen, Gorm B.},
	month = nov,
	year = {2021},
	pages = {121294},
}

@article{sasse_regional_2020,
	title = {Regional impacts of electricity system transition in {Central} {Europe} until 2035},
	volume = {11},
	copyright = {2020 The Author(s)},
	issn = {2041-1723},
	url = {https://www.nature.com/articles/s41467-020-18812-y},
	doi = {10.1038/s41467-020-18812-y},
	language = {en},
	number = {1},
	urldate = {2021-11-18},
	journal = {Nature Communications},
	author = {Sasse, Jan-Philipp and Trutnevyte, Evelina},
	month = oct,
	year = {2020},
	pages = {4972},
}

@article{pickering_diversity_2022,
	title = {Diversity of options to eliminate fossil fuels and reach carbon neutrality across the entire {European} energy system},
	volume = {6},
	issn = {2542-4785, 2542-4351},
	url = {https://www.cell.com/joule/abstract/S2542-4351(22)00236-7},
	doi = {10.1016/j.joule.2022.05.009},
	language = {English},
	number = {6},
	urldate = {2022-06-16},
	journal = {Joule},
	author = {Pickering, Bryn and Lombardi, Francesco and Pfenninger, Stefan},
	month = jun,
	year = {2022},
	note = {Publisher: Elsevier},
	pages = {1253--1276},
}

@article{victoria_speed_2022,
	title = {Speed of technological transformations required in {Europe} to achieve different climate goals},
	url = {http://arxiv.org/abs/2109.09563},
	urldate = {2022-03-24},
	journal = {arXiv:2109.09563 [physics]},
	author = {Victoria, Marta and Zeyen, Elisabeth and Brown, Tom},
	month = jan,
	year = {2022},
	note = {arXiv: 2109.09563},
}

@article{pfenninger_importance_2017,
	title = {The importance of open data and software: {Is} energy research lagging behind?},
	volume = {101},
	issn = {0301-4215},
	shorttitle = {The importance of open data and software},
	url = {https://www.sciencedirect.com/science/article/pii/S0301421516306516},
	doi = {10.1016/j.enpol.2016.11.046},
	language = {en},
	urldate = {2022-03-24},
	journal = {Energy Policy},
	author = {Pfenninger, Stefan and DeCarolis, Joseph and Hirth, Lion and Quoilin, Sylvain and Staffell, Iain},
	month = feb,
	year = {2017},
	pages = {211--215},
}

@article{pfenninger_calliope_2018,
	title = {Calliope: a multi-scale energy systems modelling framework},
	volume = {3},
	issn = {2475-9066},
	shorttitle = {Calliope},
	url = {http://joss.theoj.org/papers/10.21105/joss.00825},
	doi = {10.21105/joss.00825},
	language = {en},
	number = {29},
	urldate = {2022-03-28},
	journal = {Journal of Open Source Software},
	author = {Pfenninger, Stefan and Pickering, Bryn},
	month = sep,
	year = {2018},
	pages = {825},
}

@article{trondle_trade-offs_2020,
	title = {Trade-{Offs} between {Geographic} {Scale}, {Cost}, and {Infrastructure} {Requirements} for {Fully} {Renewable} {Electricity} in {Europe}},
	volume = {4},
	issn = {2542-4351},
	url = {https://www.sciencedirect.com/science/article/pii/S2542435120303366},
	doi = {10.1016/j.joule.2020.07.018},
	language = {en},
	number = {9},
	urldate = {2022-03-28},
	journal = {Joule},
	author = {Tröndle, Tim and Lilliestam, Johan and Marelli, Stefano and Pfenninger, Stefan},
	month = sep,
	year = {2020},
	pages = {1929--1948},
}

@article{pickering_quantifying_2021,
	title = {Quantifying resilience in energy systems with out-of-sample testing},
	volume = {285},
	issn = {0306-2619},
	url = {https://www.sciencedirect.com/science/article/pii/S0306261921000313},
	doi = {10.1016/j.apenergy.2021.116465},
	language = {en},
	urldate = {2022-06-15},
	journal = {Applied Energy},
	author = {Pickering, Bryn and Choudhary, Ruchi},
	month = mar,
	year = {2021},
	pages = {116465},
}

@article{lombardi_customised_2022,
	title = {Customised pre-built {Sector}-coupled {Euro}-{Calliope} {Model} - {Focus} on the power sector and additional {SPORES} options},
	url = {https://zenodo.org/record/6655601},
	doi = {10.5281/zenodo.6655601},
	language = {en},
	urldate = {2022-06-17},
	author = {Lombardi, Francesco},
	month = jun,
	year = {2022},
}

@article{prina_classification_2020,
	title = {Classification and challenges of bottom-up energy system models - {A} review},
	volume = {129},
	issn = {1364-0321},
	url = {https://www.sciencedirect.com/science/article/pii/S1364032120302082},
	doi = {10.1016/j.rser.2020.109917},
	language = {en},
	urldate = {2022-11-21},
	journal = {Renewable and Sustainable Energy Reviews},
	author = {Prina, Matteo Giacomo and Manzolini, Giampaolo and Moser, David and Nastasi, Benedetto and Sparber, Wolfram},
	month = sep,
	year = {2020},
	pages = {109917},
}

@misc{grochowicz_intersecting_2022,
  doi = {10.48550/ARXIV.2206.12242},
  
  url = {https://arxiv.org/abs/2206.12242},
  
  author = {Grochowicz, Aleksander and van Greevenbroek, Koen and Benth, Fred Espen and Zeyringer, Marianne},
  
  title = {Intersecting near-optimal spaces: European power systems with more resilience to weather variability},
  
  publisher = {arXiv},
  
  year = {2022},
  
  copyright = {Creative Commons Attribution 4.0 International}
}

\appendix

\renewcommand\thefigure{S\arabic{figure}}    
\setcounter{figure}{0}    

\section*{Supplementary Results}

\begin{figure}[H]
	\centerline{\includegraphics[width=1\linewidth]{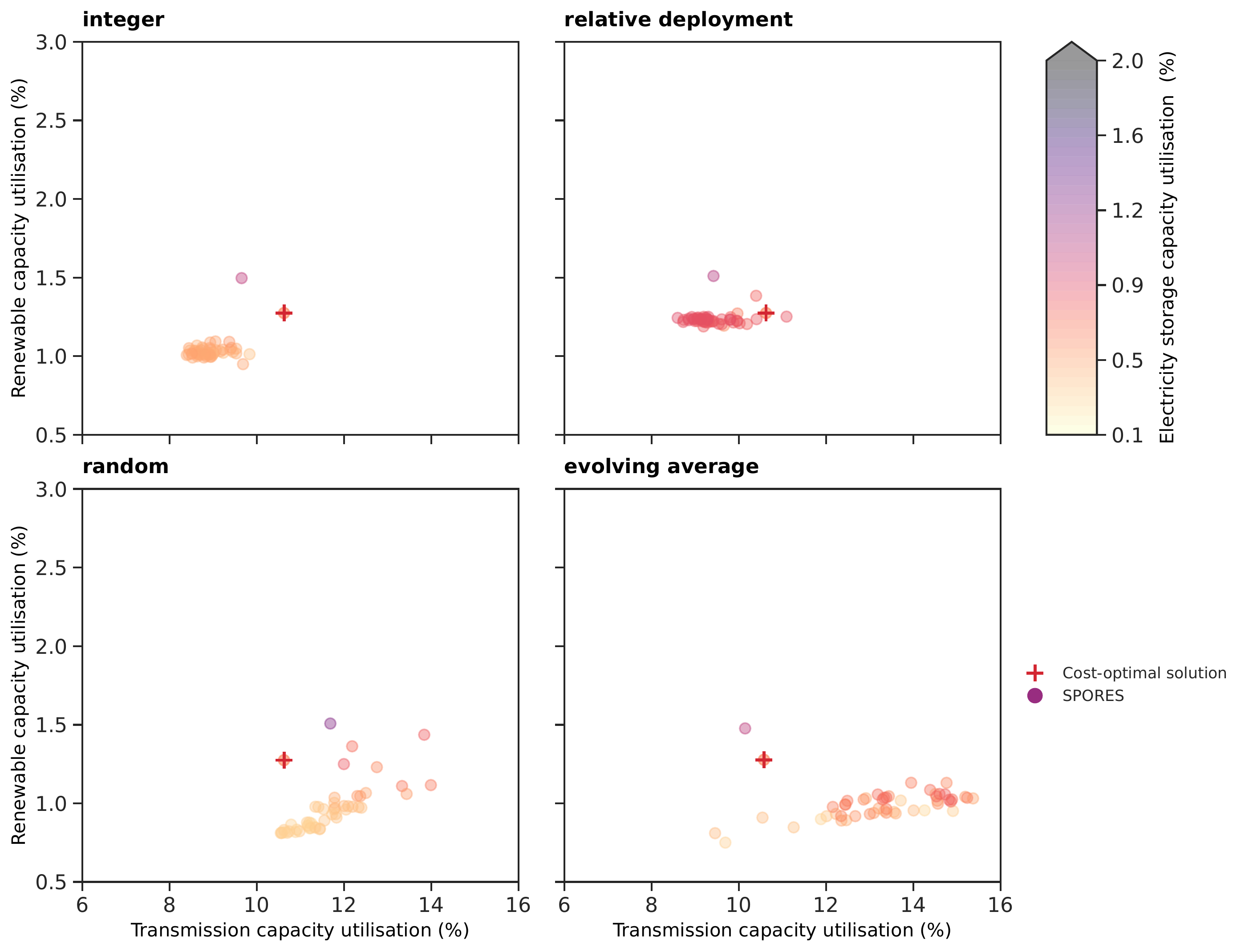}}
	\caption{Complements Figure 2 showing only the subset of 50 SPORES originating from the main batch (Equation \ref{eq:gen_spores}). Generated alternatives in a tri-dimensional space defined by aggregate renewable (Y-axis), transmission (X-axis) and storage (colour coding) capacity deployment compared to maximum deployment potential. Results are provided for all four considered search strategies. The cost-optimal solution is identified in each panel by a red marker. }
	\label{fig:sporeplot_main_batch}
\end{figure}

\begin{figure}[H]
	\centerline{\includegraphics[width=1\linewidth]{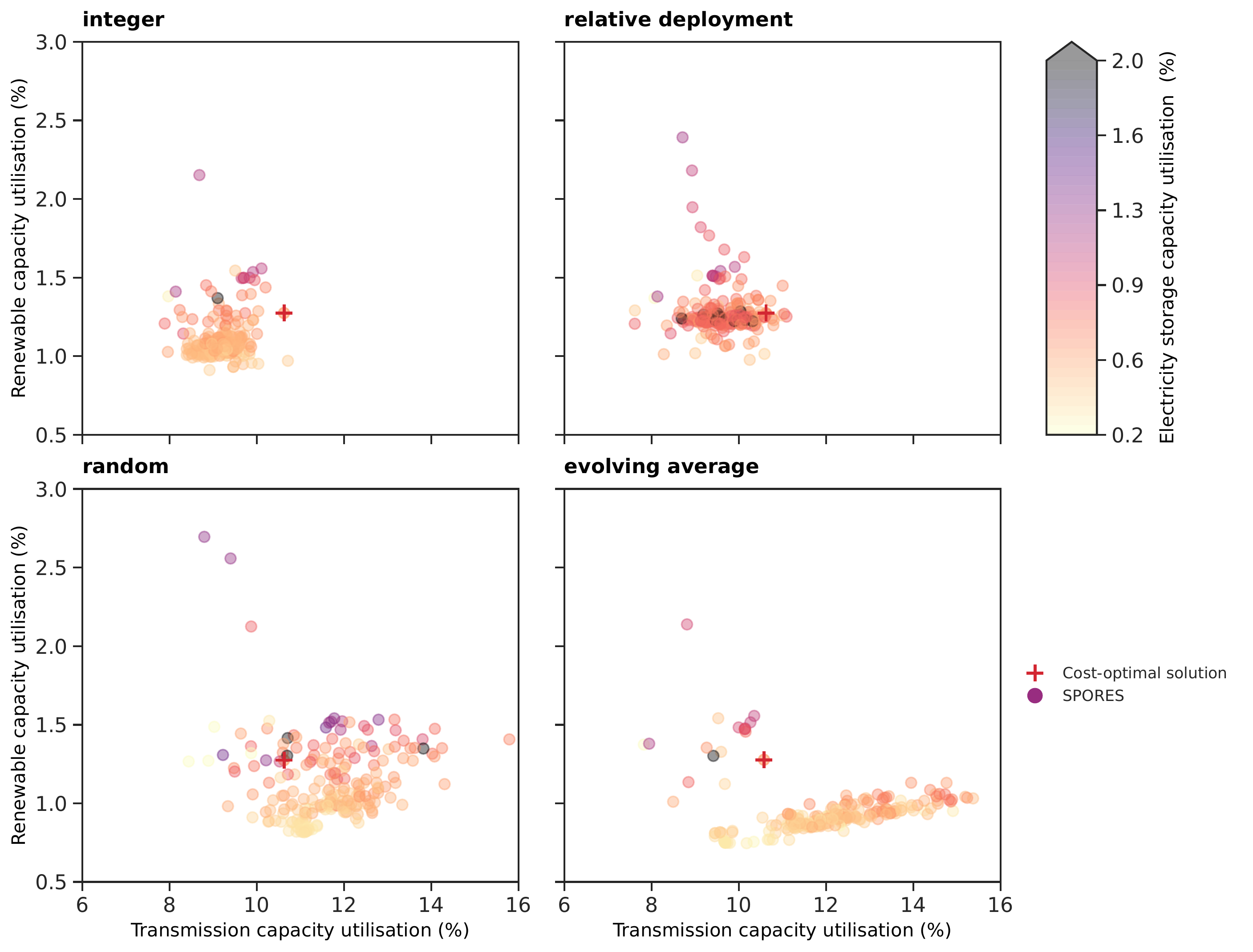}}
	\caption{Complements Figure 2 for the case of a weaker $a$ coefficient in Equation 4. Generated alternatives in a tri-dimensional space defined by aggregate renewable (Y-axis), transmission (X-axis) and storage (colour coding) capacity deployment compared to maximum deployment potential. Results are provided for all four considered search strategies. The cost-optimal solution is identified in each panel by a red marker. }
	\label{fig:sporeplot_01}
\end{figure}

\begin{figure}[H]
	\centerline{\includegraphics[width=1\linewidth]{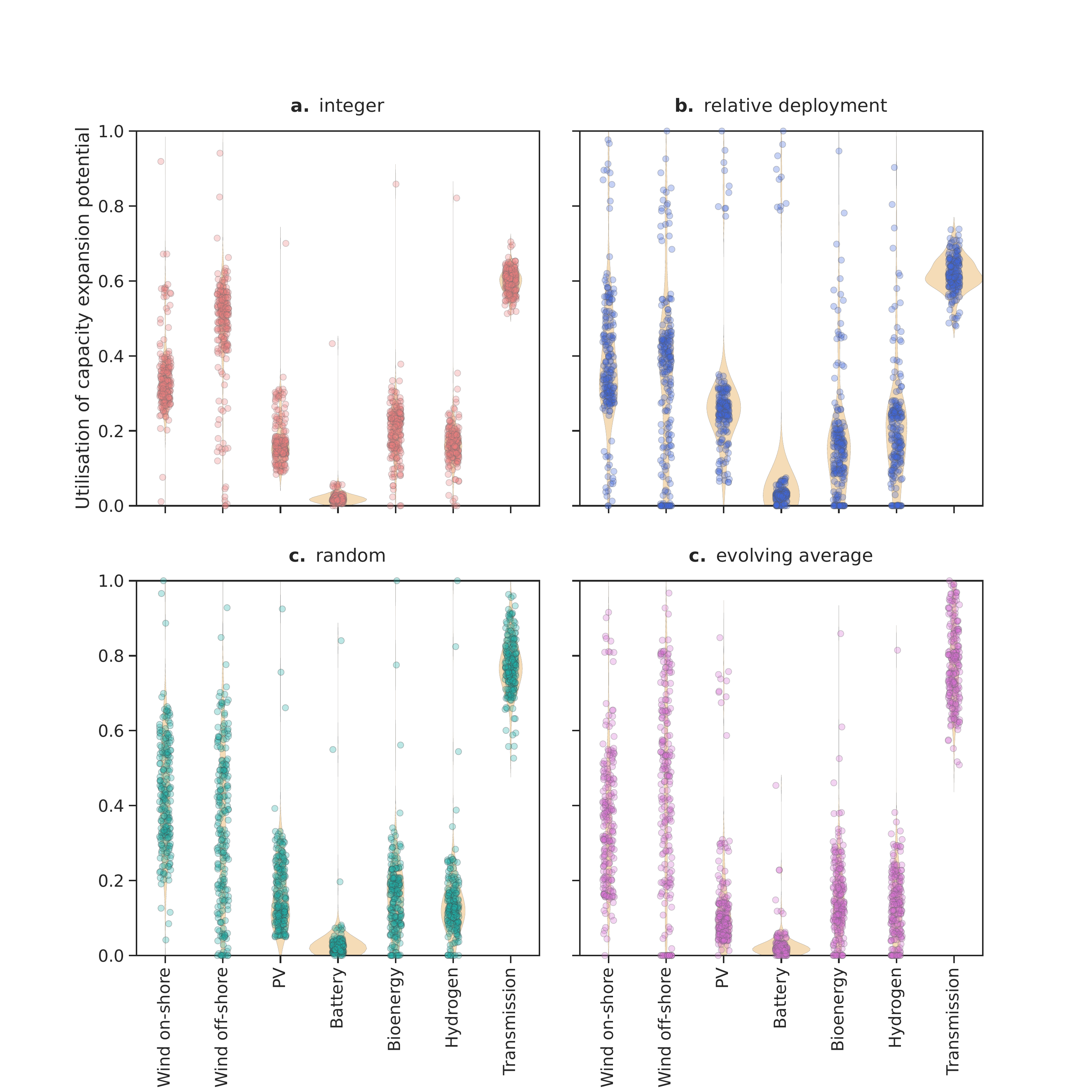}}
	\caption{Expands Figure 2 by showing the solution space generated by each search strategy from the perspective of technology-specific deployment. Generated alternatives in a one-dimensional space defined by system-wide capacity deployment for different technologies (X-axis) compared to the maximum deployment potential for each. Results are provided for all four considered search strategies. }
	\label{fig:stripviolin_1}
\end{figure}

\begin{figure}[H]
	\centerline{\includegraphics[width=1\linewidth]{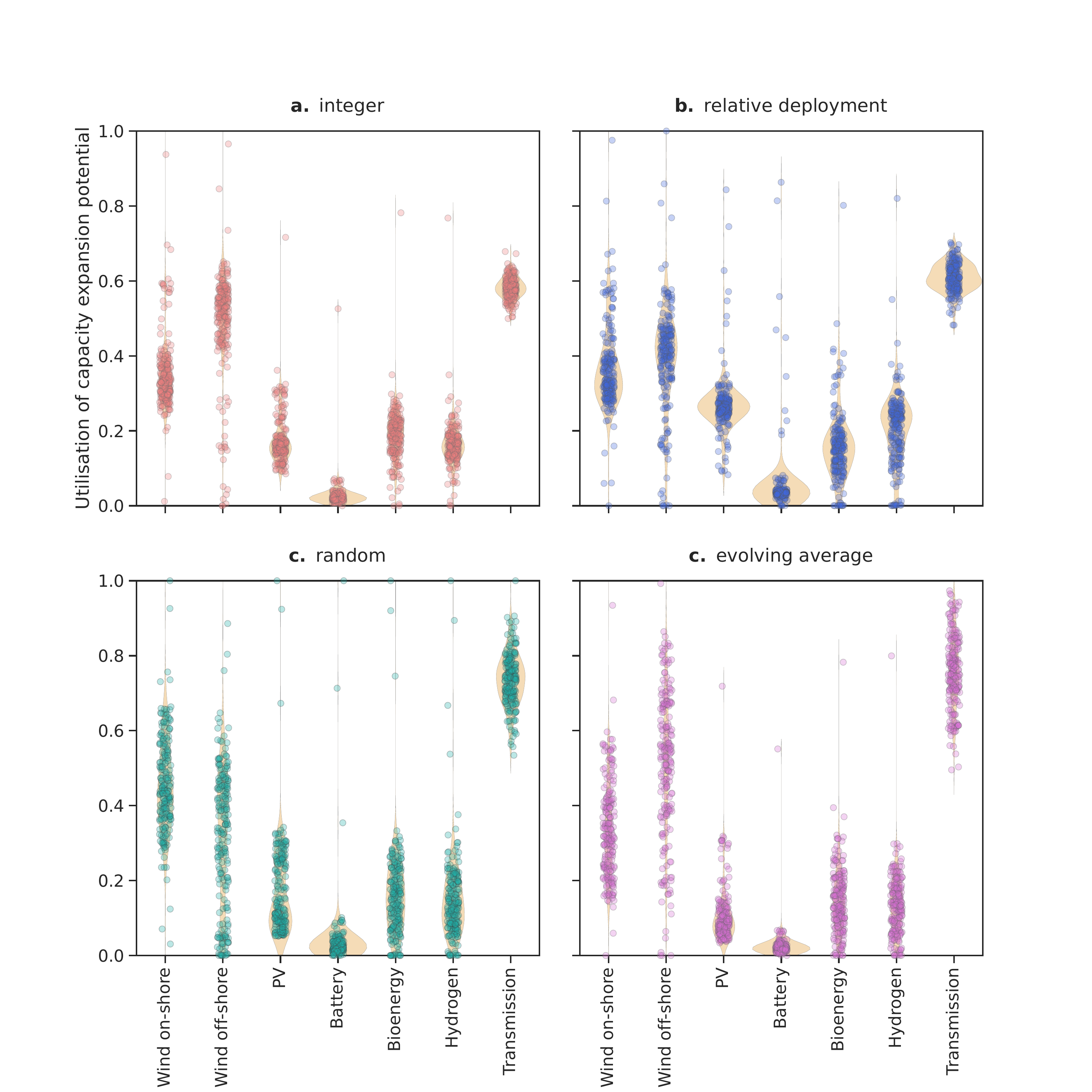}}
	\caption{Expands Figure S2 by showing the solution space generated by each search strategy, for the case of a weaker $a$ coefficient in Equation 4, from the perspective of technology-specific deployment. Generated alternatives in a one-dimensional space defined by system-wide capacity deployment for different technologies (X-axis) compared to the maximum deployment potential for each. Results are provided for all four considered search strategies. }
	\label{fig:stripviolin_01}
\end{figure}

\begin{figure}[H]
	\centering
	\includegraphics[width=1\linewidth]{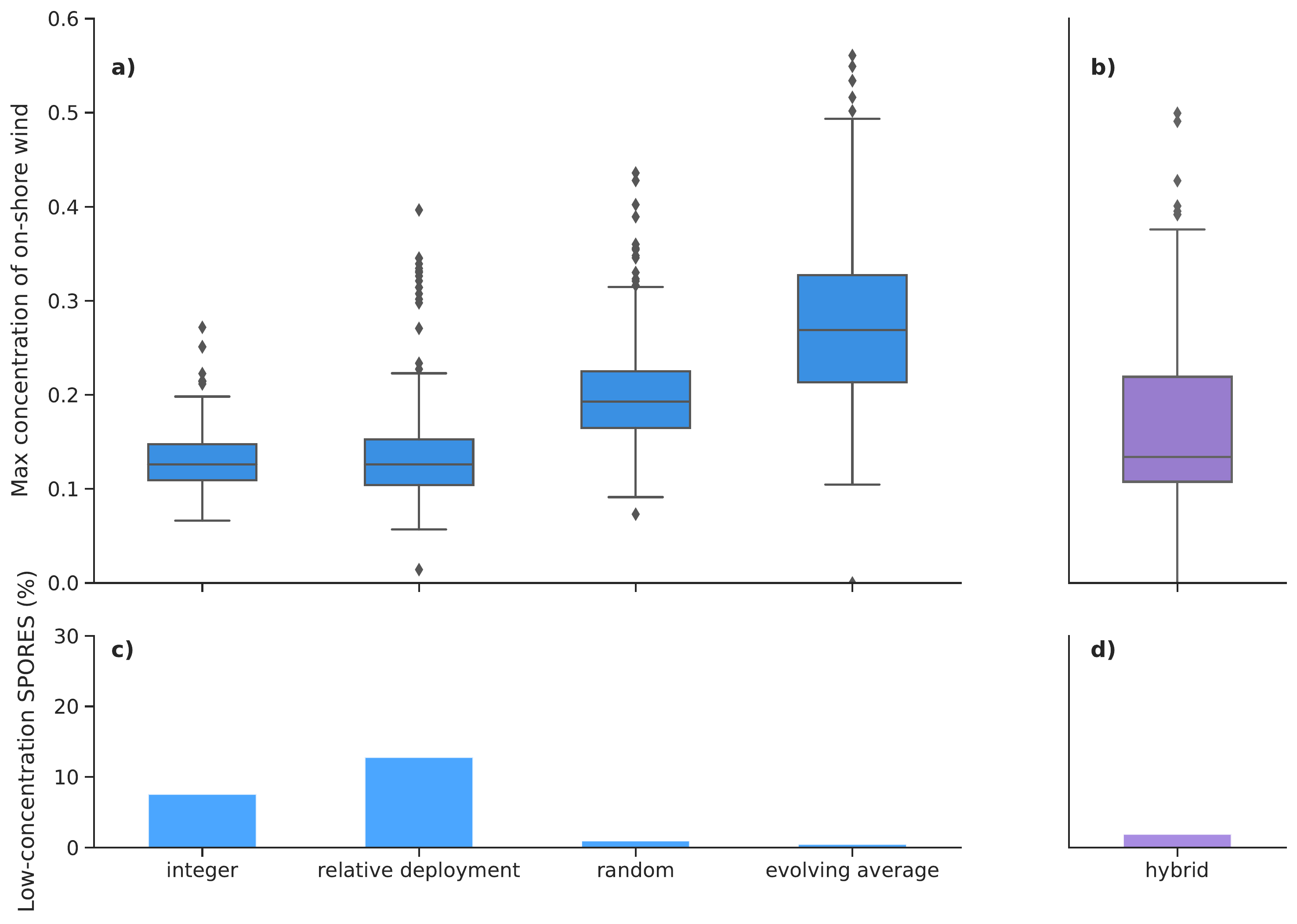}
	\caption{Complements Figure 3 for the case of a weaker $a$ coefficient in Equation 4. The top of the Figure shows the data distribution (boxplots) for the maximum concentration of deployed on-shore wind capacity in a single region (across all the model 97 locations) compared to the total deployed on-shore wind capacity. The boxplots are shown for the four tested weight-assignment methods (panel \textbf{a}) and for a hybrid case that combines ex-post the data from the relative-deployment and evolving-average methods (panel \textbf{b}); see subsection \ref{results-useful-spores} for further details about such a hybrid case. The bottom of the Figure shows the absolute number of generated alternatives that have a concentration of on-shore wind capacity at least 20\% lower than in the cost-optimal case. Results are again shown for the four tested weight-assignment methods (panel \textbf{c}) and for the hybrid case (\textbf{d}). }
	\label{fig:boxplots_01}
\end{figure}

\pagebreak

\section*{Supplementary Methods} \label{supp_methods}

The model we use in this study is based on the open-source Sector-Coupled Euro-Calliope modelling workflow (\url{https://github.com/calliope-project/sector-coupled-euro-calliope}). More precisely, we customise a pre-built Sector-Coupled Euro-Calliope model to focus only on the power sector. The main customisations we apply here are the following:

\begin{itemize}
    \item We do not load any model scenario related to sectors other than power.
    \item We modify the electricity demand time-series file, originally net of the electricity consumed in other sectors, to include also the electricity that is currently consumed for heat, mobility or industrial energy services.
    \item We eliminate Iceland from the analysis since we do not model geothermal energy, which accounts for about 30\% of Iceland's electricity production today. This does not have much of an impact on results, due to the small size of the country and its distance from the rest of the continent. 
    \item We introduce new, ad-hoc SPORES constraints in the model, which require a specific Calliope environment to be run. We include the requirement and indications about how to meet them in our public release of the code.
\end{itemize}

We make the code openly available on Zenodo (\url{https://doi.org/10.5281/zenodo.6655601}), including brief documentation about how to use it. 

In addition to the information provided in sub-section \ref{methods-eurocalliope}, further details about the Euro-Calliope model that we use as a starting point, including technology cost assumptions, are available from the associated publication \cite{pickering_diversity_2022}. Nonetheless, we report in the sub-section below the key features and assumptions of our customised model setup to improve its understandability, particularly for readers that are less familiar with the class of models in question.

\subsection*{Key features and assumptions of the model setup for this study}

\begin{itemize}
    \item \textbf{Model formulation:} linear programming (LP) generation of alternative feasible system designs starting from the least-cost feasible solution. The generation of alternative solutions is based on the SPORES algorithm, an original advancement of MGA. 
    \item \textbf{Geographical scope:} 34 European countries, 97 sub-regions.
    \item \textbf{Temporal scope:} one representative year of weather and demand data (2018), 3-hour resolution. The system is designed considering the lifetime of technologies and their annualised cost. There is no analysis of system evolution through time to reach the end-state design.
    \item \textbf{Costs:} projections considering a 2050 desired end state and an interest rate of 0.073. Detailed sources are provided for each technology and for the interest rate as comments in the associated input files, available on Zenodo. 
    \item \textbf{Boundary conditions:} no net CO$_2$ emissions. Bioenergy is considered carbon-neutral.
    \item \textbf{Aspects beyond the model scope:} 
        \begin{itemize}
            \item \textit{Technologies:} no carbon capture and storage, limited allowed deployment of nuclear energy (aligned to current policies of each country)
            \item \textit{Sectors:} no energy sectors outside the power system. The electricity demand does include currently electrified heat, mobility and industry uses of energy.
            \item \textit{Phenomena:} no frequency and voltage regulation of the grid.
        \end{itemize} 
\end{itemize}

\end{document}